\documentclass[a4]{jpconf}
\usepackage{graphicx}
\begin{document}
\title{Evidence for power-law Griffiths singularities in a layered Heisenberg magnet}

\author{Fawaz Hrahsheh$^1$, Hatem Barghathi$^1$, Priyanka Mohan$^2$, Rajesh Narayanan$^2$ and Thomas Vojta$^1$}

\address{$^1$ Department of Physics, Missouri University of Science and Technology, Rolla MO 65409, USA}
\address{$^2$ Department of Physics, Indian Institute of Technology Madras, Chennai 600036, India}

\ead{vojtat@mst.edu}

\begin{abstract}
We study the ferromagnetic phase transition in a randomly layered Heisenberg model.
A recent strong-disorder renormalization group approach [Phys.\ Rev.\ B {\bf 81},
144407 (2010)] predicted that the critical
point in this system is of exotic infinite-randomness type and is accompanied by
strong power-law Griffiths singularities. Here, we report results of Monte-Carlo
simulations that provide numerical evidence in support of these predictions.
Specifically, we investigate the finite-size scaling behavior of the magnetic
susceptibility which is characterized by a non-universal power-law divergence in
the Griffiths phase. In addition, we calculate the time autocorrelation function
of the spins.
It features a very slow decay in the Griffiths phase, following a non-universal
power law in time.
\end{abstract}

\section{Introduction}

Impurities, defects, and other types of quenched disorder influence zero-temperature
quantum phase transitions much more strongly than generic classical phase transitions
that occur at nonzero temperatures. The interplay of quantum and disorder fluctuations
produces uncon\-ventional phenomena such as power-law quantum Griffiths singularities
\cite{ThillHuse95,GuoBhattHuse96,RiegerYoung96}, infinite-randomness critical points
featuring exponential instead of power-law scaling \cite{Fisher92,Fisher95},
and smeared phase transitions \cite{Vojta03a,HoyosVojta08}. A recent review of
these phenomena can be found in Ref.\ \cite{Vojta06}, while Ref.\ \cite{Vojta10}
focuses on metallic systems and also discusses experiments.

Disorder effects are stronger at quantum phase transitions than at
classical transitions because the quenched disorder is perfectly correlated in
the \emph{imaginary time} direction. Imaginary time acts as an extra dimension at a
quantum phase transition and becomes infinitely extended at zero temperature.
Therefore, the impurities and defects are effectively ``infinitely  large'' in this extra dimension,
which makes them much harder to average out than finite-size defects.

This argument suggests that strong disorder effects should also occur at a classical
thermal phase transition provided that the disorder is perfectly correlated in one or
more \emph{space} dimensions. An example occurs in the McCoy-Wu model, a
disordered classical two-dimensional Ising model having perfect disorder correlations
in one of the two dimensions. McCoy and Wu \cite{McCoyWu68,McCoyWu68a,McCoyWu69,McCoy69}
showed that this model exhibits an unusual phase transition featuring a smooth specific heat
while the susceptibility is infinite over an entire temperature range.
Fisher \cite{Fisher92,Fisher95} achieved an essentially complete understanding of
this phase transition with the help of a strong-disorder renormalization group approach
(using the equivalence between the McCoy-Wu model and the random transverse-field Ising
chain). He determined that the critical point is of exotic infinite-randomness type
and is accompanied by power-law Griffiths singularities.

Recently, some of us investigated another classical system with perfectly correlated
disorder, the randomly layered Heisenberg magnet, by means of a strong-disorder
renormalization group \cite{MohanNarayananVojta10}. This theory predicts that the
(thermal) critical point in the randomly layered Heisenberg magnet is of infinite-randomness
type as well. Moreover, it is in the same universality class as the quantum critical point
of the random transverse-field Ising chain.
In this paper, we present the results of Monte-Carlo simulations of the randomly
layered Heisenberg model. They provide numerical evidence in support of the above
renormalization group predictions.

\section{Model and renormalization group predictions}
\label{sec:RG}

We consider a ferromagnet consisting of a random sequence of
layers made up of two different ferromagnetic materials. Its Hamiltonian, a classical Heisenberg
model on a three-dimensional lattice of perpendicular size $L_\perp$ (in $z$ direction) and
in-plane size $L_\parallel$ (in the $x$ and $y$ directions)  is given by
\begin{equation}
H = - \sum_{\mathbf{r}} J^{\parallel}_z \, (\mathbf{S}_{\mathbf{r}} \cdot \mathbf{S}_{\mathbf{r}+\hat{\mathbf{x}}}
                                        +\mathbf{S}_{\mathbf{r}} \cdot \mathbf{S}_{\mathbf{r}+\hat{\mathbf{y}}} )
    - \sum_{\mathbf{r}} J^{\perp}_z \, \mathbf{S}_{\mathbf{r}} \cdot\mathbf{S}_{\mathbf{r}+\hat{\mathbf{z}}}
    .
\label{Eq:Hamiltonian}
\end{equation}
Here, $\mathbf{S}_{\mathbf{r}}$ is a three-component unit vector on lattice site
$\mathbf{r}$, and  $\hat{\mathbf{x}}$, $\hat{\mathbf{y}}$, and $\hat{\mathbf{z}}$ are
the unit vectors in the coordinate directions. The interactions within
the layers, $J^{\parallel}_z$, and between the layers, $J^{\perp}_z$, are both positive
and independent random functions of the perpendicular coordinate $z$. In the following,
we take all $J^{\perp}_z$ to be identical, $J^{\perp}_z \equiv J^{\perp}$,
while the $J^{\parallel}_z$ are drawn from a binary probability distribution
$P(J^{\parallel})=(1-p)\, \delta(J^{\parallel} - J_u) + p\, \delta(J^{\parallel} - J_l)$
with $J_u > J_l$. Here, $p$ is the concentration of the ``weak'' layers.

The qualitative behavior of (\ref{Eq:Hamiltonian}) is easily explained. At sufficiently high
temperatures, the model is in a conventional paramagnetic
phase. Below a temperature $T_u$ (the transition temperature of a hypothetical system having
$J^{\parallel}_z \equiv J_u$ for all $z$) but above the actual critical temperature $T_c$,
rare thick slabs of strong layers develop local order while the bulk system is still
nonmagnetic. This is the \emph{paramagnetic} Griffiths phase (or Griffiths region). In the \emph{ferromagnetic} Griffiths phase,
located between $T_c$ and a temperature $T_l$ (the transition temperature of a hypothetical
system having $J^{\parallel}_z \equiv J_l$ for all $z$), bulk magnetism coexists with rare
nonmagnetic slabs.  Finally, below
$T_l$, the system is in a conventional ferromagnetic phase.

In Ref.\ \cite{MohanNarayananVojta10}, the behavior in both Griffiths phases and at criticality
has been derived within a strong-disorder renormalization group calculation. Here, we simply
motivate and summarize the results. The probability of finding a slab of
$L_{RR}$ consecutive strong layers is given by simple combinatorics; it reads
$w(L_{RR}) \sim (1-p)^{L_{RR}} = e^{-\tilde p L_{RR} }$
with $\tilde p = -\ln(1-p)$. Each such slab is equivalent to a two-dimensional
Heisenberg model with an effective interaction $L_{RR} J_u$. Because the two-dimensional
Heisenberg model is exactly at its lower critical dimension, the renormalized
distance from criticality, $\epsilon$, of such a slab decreases exponentially with its thickness,
$\epsilon(L_{RR}) \sim e^{-b L_{RR}}$
\cite{Vojta06,VojtaSchmalian05}.
Combining the two exponentials gives a power law spectrum of locally ordered slabs,
\begin{equation}
\rho(\epsilon) \sim \epsilon^{\tilde p / a -1} = \epsilon^{1/z-1}
\label{Eq:DOS}
\end{equation}
where the second equality defines the conventionally used dynamical exponent, $z$.
It increases with decreasing temperature throughout the Griffiths phase and
diverges as $z \sim 1/|T-T_c|$ at the actual critical point.

Many important observables follow from appropriate integrals of (\ref{Eq:DOS}).
The susceptibility can be estimated by $\chi \sim \int d\epsilon\, \rho(\epsilon)/\epsilon$.
In an infinite system, the lower bound of the integral is 0; therefore, the susceptibility
diverges in the entire temperature region where $z>1$. A finite system size $L_\parallel$
in the in-plane directions introduces a nonzero lower bound $\epsilon_{\rm min} \sim
L_\parallel^{-2}$. Thus, the susceptibility in the Griffiths region diverges as
\begin{equation}
\chi (L_\parallel) \sim L_\parallel^{2-2/z}
\label{Eq:chi(Lt)}
\end{equation}
for $z>1$. The behavior of the time autocorrelation function $C(t) = (1/N) \sum_{\mathbf{r}} \langle
\mathbf S (\mathbf{r},t)\cdot \mathbf S(\mathbf{r},0)\rangle$ within model A dynamics
\cite{HohenbergHalperin77} can be determined analogously ($N$ is the total number of spins).
As the correlation time of a single, locally ordered slab is inversely proportional
to its renormalized distance from criticality, we obtain, in the Griffiths phase,
\begin{equation}
C(t) \sim \int d\epsilon\, \rho(\epsilon) e^{-\epsilon t} \sim t^{-1/z}~.
\label{Eq:C(t)}
\end{equation}

\section{Monte-Carlo simulations}
\label{sec:MC}

We have carried out Monte-Carlo simulations of the Hamiltonian (\ref{Eq:Hamiltonian})
with system sizes up to $L_\perp=800$ and $L_\parallel=150$. While studying the thermodynamics,
we have used the efficient Wolff cluster algorithm \cite{Wolff89} to eliminate critical slowing down.
To measure the time autocorrelation function, we have equilibrated the system using the Wolff algorithm but
then propagated the dynamics by means of the Metropolis algorithm \cite{MRRT53} which implements model A
dynamics. All results reported below are for $J_z^\perp \equiv 1.00$ and a concentration $p=0.8$ of weak layers with $J_u =1.00$ and
$J_l=0.25$. The data are averages over a large number (20 to 160) of disorder realizations.

\begin{figure}
\includegraphics[width=7cm]{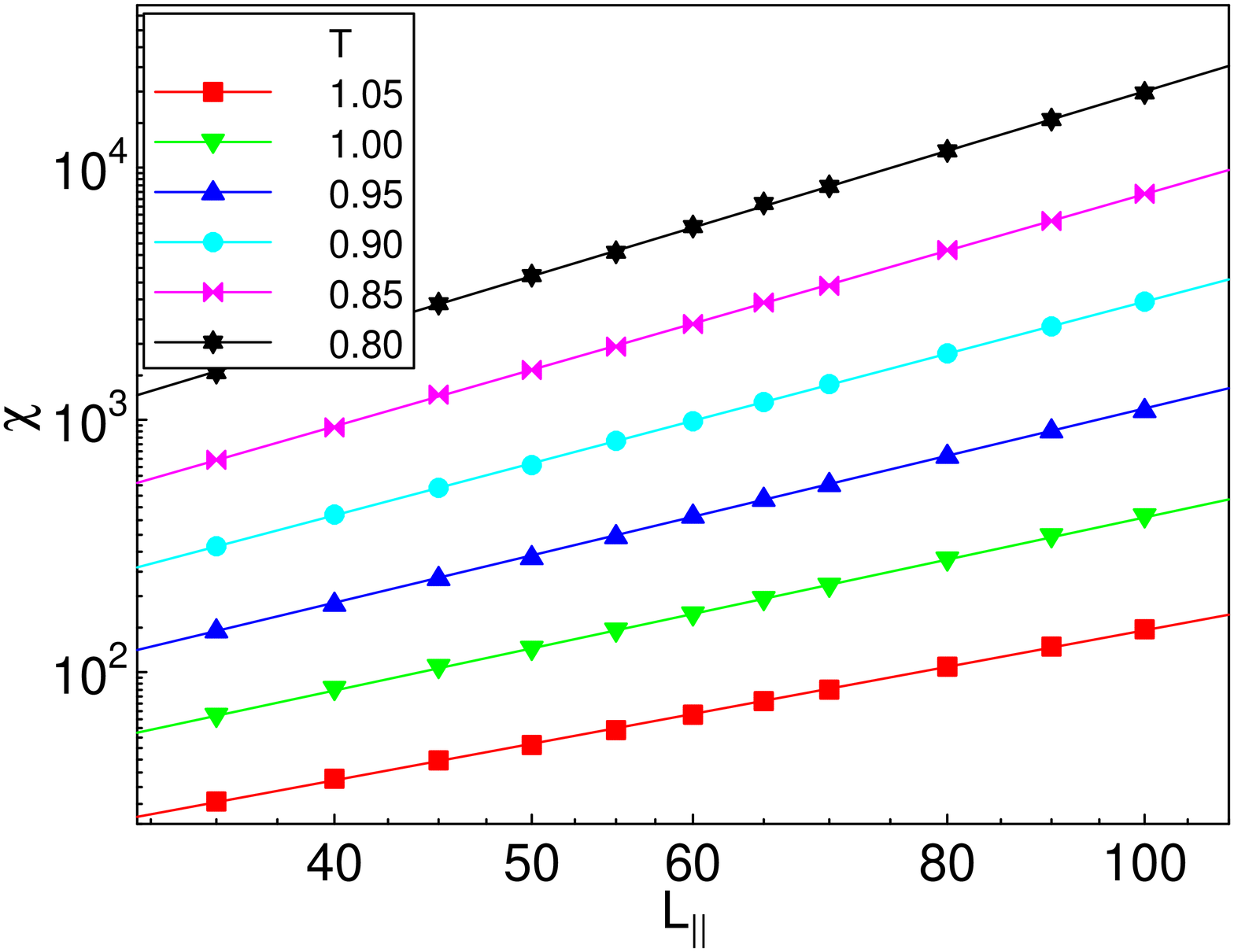} \hfill \includegraphics[width=7cm]{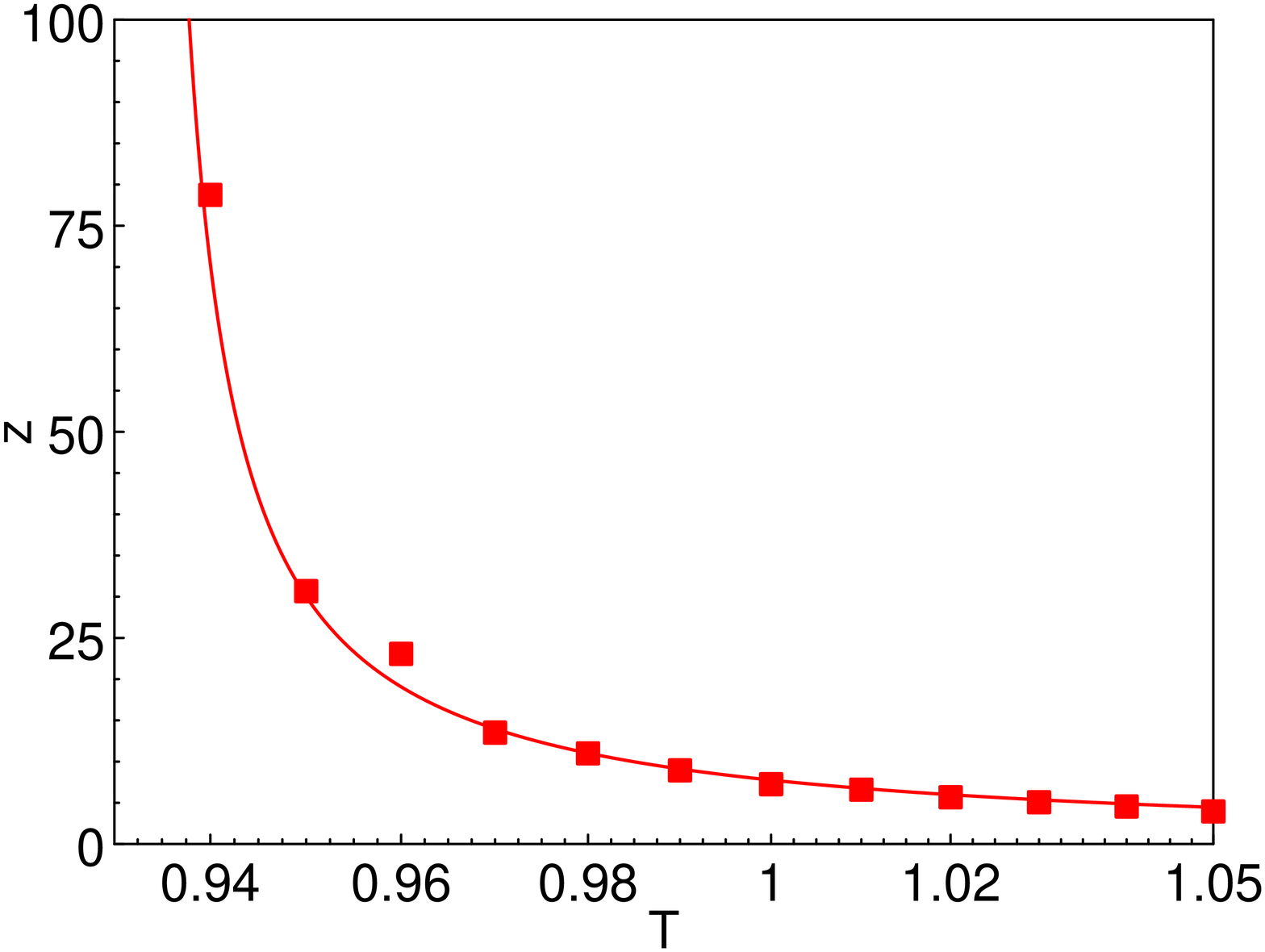}
\caption{\label{Fig:1}Left: Susceptibility $\chi$ as a function of in-plane system size $L_\parallel$
         for several temperatures in the Griffiths region. The perpendicular size is $L_\perp=800$;
         the data are averages over 160 disorder configurations. The solid lines are fits
         to the power law (\ref{Eq:chi(Lt)}). Right: Resulting Griffiths dynamical exponent $z$
         as a function of temperature $T$ in the weakly disordered Griffiths region.}
\end{figure}

To test the finite-size behavior (\ref{Eq:chi(Lt)}) of the susceptibility, one needs to consider samples having sizes
$L_\perp \gg L_\parallel$ such that $L_\perp$ is effectively infinite. We have used system sizes
$L_\perp=800$ and $L_\parallel=35$ to 100. Figure \ref{Fig:1} shows the susceptibility $\chi$ as a function of $L_\parallel$
for several temperatures in the Griffiths region below $T_u \approx 1.443$.
$\chi$ follows a nonuniversal power law
in  $L_\parallel$ with a temperature-dependent exponent. Simulations for many more
temperature values yield analogous results. The value of the exponent $z$ extracted from fits to
(\ref{Eq:chi(Lt)}) is shown in the right panel of figure \ref{Fig:1} for the paramagnetic side of
the Griffiths region. $z$ can be fitted to the predicted power law $z\sim 1/|T-T_c|$, as discussed
after (\ref{Eq:DOS}), giving the estimate $T_c \approx 0.933$

\begin{figure}
\includegraphics[width=7cm]{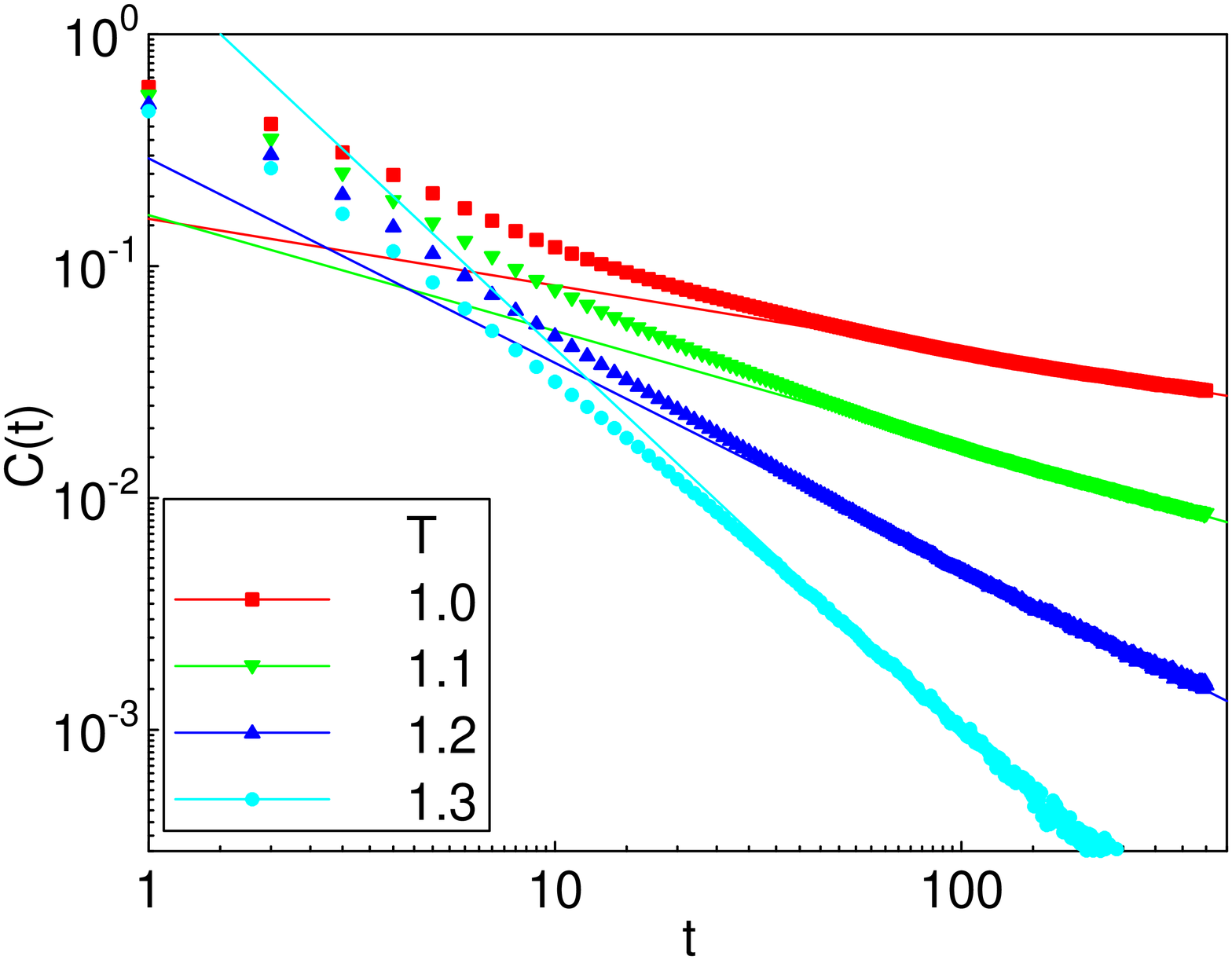} \hfill\begin{minipage}[b]{7cm}
\caption{\label{Fig:2}Time autocorrelation function $C(t)$ for several temperatures in the Griffiths phase.
         The system sizes are $L_\perp=400$ and $L_\parallel=100$. The data are averages over
         80 disorder configurations. The solid lines are fits to (\ref{Eq:C(t)}).\\ }
\end{minipage}
\end{figure}

Figure \ref{Fig:2}
shows the autocorrelation function $C(t)$ of the spins as a function of time $t$ for several temperatures in the Griffiths phase.
In agreement with the prediction (\ref{Eq:C(t)}), the long-time behavior of $C(t)$ follows
a nonuniversal power law. We have also studied (not shown) the behavior at criticality. It displays
the logarithmic time-dependence predicted in Ref.\ \cite{MohanNarayananVojta10}.

\section{Conclusions}
\label{sec:Conclusions}
In summary, we have performed Monte-Carlo simulations of a randomly layered three-dimensional
Heisenberg model. Our numerical results for the magnetic susceptibility and the time autocorrelation function
in the paramagnetic Griffiths phase  support the infinite-randomness critical point scenario
 that arises from the strong-disorder renormalization group approach \cite{MohanNarayananVojta10}. A full confirmation
of the theoretical predictions requires performing a scaling analysis of the infinite-randomness critical point itself,
and measuring all three independent critical exponents. This work is in progress.

Experimental verifications of infinite-randomness critical behavior and the
accompanying power-law Griffiths singularities have been hard to come
by, in particular in higher-dimensional systems. Only very recently, promising
measurements have been reported \cite{Westerkampetal09,UbaidKassisVojtaSchroeder10}
of the quantum phase transitions in CePd$_{1-x}$Rh$_x$ and Ni$_{1-x}$V$_x$.
The randomly layered Heisenberg magnet considered here provides an alternative realization
of an infinite-randomness critical point.  It may be more easily realizable in experiment
because the critical point is classical, and samples can be produced by depositing
random layers of two different ferromagnetic materials.


This work has been supported in part by the NSF under grant nos. DMR-0339147
and DMR-0906566 and by Research Corporation.

\section*{References}

\bibliographystyle{iopart-num}
\bibliography{../../00Bibtex/rareregions}

\end{document}